\documentclass[a4paper,twoside,12pt]{article}
\input{obsjkt.sty}

\newcommand{\Msun}{\ensuremath{~{\rm M}_\odot}}                   % Solar mass symbol
\newcommand{\Rsun}{\ensuremath{~{\rm R}_\odot}}                   % Solar radius symbol
\newcommand{\rhosun}{\ensuremath{~\rho_\odot}}                    % Solar density symbol
\newcommand{\Teff}{\ensuremath{T_{\rm eff}}}                      % Effective temperature symbol
\newcommand{\EBV}{\ensuremath{E(B\!-\!V)}}                        % E(B-V) symbol
\newcommand{\degr}{\ensuremath{^\circ}}                           % Degree symbol
\renewcommand{\kms}{~km~s$^{-1}$}                                 % km/s symbol
\renewcommand{\cd}{~d$^{-1}$}                                     % Cycles per day symbol
\newcommand{\chisq}{\ensuremath{\chi^{\,2}}}                      % Chi-squared symbol
\newcommand{\mc}[1]{\multicolumn{2}{c}{#1}}                       % centred between two columns
\newcommand{\gaia}{\textit{Gaia}}                                 % Gaia mission
\newcommand{\targ}{IQ~Per}
\newcommand{\targfull}{IQ~Persei}

\newcommand{\Msunnom}{\hbox{$\mathcal{M}^{\rm N}_\odot$}}
\newcommand{\Rsunnom}{\hbox{$\mathcal{R}^{\rm N}_\odot$}}
\newcommand{\Lsunnom}{\hbox{$\mathcal{L}^{\rm N}_\odot$}}

\usepackage{rotating}

\begin{document} %%%%%%%%%%%%%%%%%%%%%%%%%%%%%%%%%%%%%%%%%%%%%%%%%%%%%%%%%%%%%%%%%%%%%%%%%%%%%%%%%%%%%%%%%%%%%%%%%%%%%%%%%%%%%%%%%%%%%%%%%%%%%%%%%%%%
%%%%%%%%%%%%%%%%%%%%%%%%%%%%%%%%%%%%%%%%%%%%%%%%%%%%%%%%%%%%%%%%%%%%%%%%%%%%%%%%%%%%%%%%%%%%%%%%%%%%%%%%%%%%%%%%%%%%%%%%%%%%%%%%%%%%%%%%%%%%%%%%%%%%%

\OBSheader{Rediscussion of eclipsing binaries: \targ}{J.\ Southworth}{2024 December}

\OBStitle{Rediscussion of eclipsing binaries. Paper XXI. \\ The totally-eclipsing B-type system IQ Persei}

\OBSauth{John Southworth}

\OBSinstone{Astrophysics Group, Keele University, Staffordshire, ST5 5BG, UK}

%\OBSabstract{Abs abs abs abs abs abs abs abs abs abs abs abs abs abs abs abs abs abs abs abs abs abs abs abs abs abs abs abs abs abs abs abs abs abs abs abs abs abs abs abs abs abs abs abs abs abs abs abs abs abs abs abs abs abs abs abs abs abs abs abs abs abs abs abs abs abs abs abs abs abs abs abs abs abs abs abs abs abs abs abs abs abs abs abs abs abs abs abs abs abs abs abs abs abs abs abs abs abs abs abs abs abs abs abs abs abs abs abs abs abs abs abs abs abs abs abs abs abs abs abs abs abs abs abs abs abs abs abs abs abs abs abs abs abs abs abs abs abs abs abs abs abs abs abs abs abs abs abs abs abs abs abs abs abs abs abs abs abs abs abs abs abs abs abs abs abs abs abs abs abs abs abs abs abs abs abs abs abs abs abs abs abs abs abs abs abs abs abs abs abs abs abs abs abs abs abs abs abs abs abs abs abs abs abs abs abs abs abs abs abs abs abs abs abs abs abs abs abs abs abs abs abs abs abs abs abs abs abs abs abs abs abs abs abs abs abs abs abs abs abs abs abs abs.}

\OBSabstract{\targ\ is a totally-eclipsing binary system containing a B8~V star and an A6~V star in an orbit of period 1.744~d with eccentricity and apsidal motion. We use new light curves from the Transiting Exoplanet Survey Satellite (TESS) and published spectroscopy from Lacy \& Frueh \cite{LacyFrueh85apj} to measure the physical properties of the component stars, finding masses of $3.516 \pm 0.050$\Msun\ and $1.738 \pm 0.023$\Msun, and radii of $2.476 \pm 0.015$\Rsun\ and $1.503 \pm 0.016$\Rsun. Our fit to the light curve is imperfect, with a small sinusoidal trend in the residuals versus orbital phase and a slight mismatch in the depth of secondary eclipse, but the total eclipses mean the system is still well-characterised. The distance to the system from its masses, temperatures, apparent magnitudes and bolometric corrections is in agreement with the parallax distance from \gaia\ DR3. Theoretical models cannot adequately match the measured properties of the system, and new spectroscopy to confirm the temperatures and determine the chemical compositions of the stars would be useful. A Fourier analysis of the residuals of the best fit to the light curve shows many peaks at multiples of the orbital frequency, and one significant peak at 1.33\cd\ which is not. This pulsation and the properties of the primary component are consistent with it being a slowly-pulsating B star.}

%%%%%%%%%%%%%%%%%%%%%%%%%%%%%%%%%%%%%%%%%%%%%%%%%%%%%%%%%%%%%%%%%%%%%%%%%%%%%%%%%%%%%%%%%%%%%%%%%%%%%%%%%%%%%%%%%%%%%%%%%%%%%%%%%%%%%%%%%%%%%%%%%%%%%

\section*{Introduction}

This work continues our series of papers \cite{Me20obs} presenting analyses of detached eclipsing binaries (dEBs) with a significant observational history and available radial velocity (RV) measurements, based on new high-quality light curves from the Transiting Exoplanet Survey Satellite (TESS \cite{Ricker+15jatis}). Our aim is to increase the number of stars, and the precision of their measured properties, in the Detached Eclipsing Binary Catalogue \cite{Me15aspc} (DEBCat\footnote{\texttt{https://www.astro.keele.ac.uk/jkt/debcat/}}), which lists all known dEBs with mass and radius measurements to 2\% precision and accuracy. These results represent an important resource of empirical measurements of stellar properties against which theoretical models of stellar evolution can be benchmarked \cite{Andersen91aarv,Torres++10aarv}, and the availability of large archives of light curves from space missions enables such work for a large number of objects \cite{Me21univ}.

Here we study the system \targfull\ (Table~\ref{tab:info}), which contains a late-B star (hereafter star~A) and an early-A star (star~B) in an orbit of period 1.744~d. It is one of many whose variability was discovered by Hoffmeister \cite{Hoffmeister49book} using photographic plates obtained at Sonneberg. It is a visual double with a companion at 39.3~arcsec which is fainter by 1.58~mag in the \gaia\ $G$ band. Meisel \cite{Meisel68aj} assigned spectral types of B8~Vp: to \targ\ and A0~Vnp: to the visual companion. Burke \cite{Burke68ibvs} narrowed down the orbital period to be either 6.974~d or its submultiples 3.487~d or 1.743~d. Hall, Gertken \& Burke \cite{Hall++70pasp} presented $UBV$ light curves which confirmed the shortest of the possible orbital periods, the eccentricity, and that the system exhibits total eclipses. They also derived its photometric properties using the Russell-Merrill \cite{RussellMerrill52book} method. Bischof \cite{Bischof72ibvs} presented new times of minimum and Young \cite{Young75pasp} a first spectroscopic orbit for both stars.

Lacy \& Frueh \cite{LacyFrueh85apj} (hereafter LF85) published a detailed analysis of \targ, and their measurements of the physical properties have been used in many subsequent papers. They obtained a set of 20 spectra using a Reticon detector at the 2.7~m telescope at McDonald Observatory, measuring from these RVs of both stars plus projected rotational velocities of $V_{\rm A} \sin i = 68 \pm 2$\kms\ and $V_{\rm B} \sin i = 44 \pm 2$\kms, where `A' denotes the more massive primary (star~A) and `B' the less massive secondary (star~B). LF85 also obtained light curves in the $V$ and $R$ bands; these data cover all of the secondary and almost all of the primary eclipse. They deduced photometric spectral types of B8 for star~A and A6 for star~B.

\targ\ also shows apsidal motion dominated by tidal effects, and this has been measured by a multitude of researchers using essentially the same gradually-growing compilation of times of mid-eclipse. Apsidal motion was predicted by Hall et al.\ \cite{Hall++70pasp}, and confirmed by LF85 who found an apsidal period of $U = 140 \pm 30$~yr. Drozdz et al.\ \cite{Drozdz++90ibvs} improved this measurement to $119 \pm 9$~yr. De{\u{g}}irmenci \cite{Degirmenci97apss} obtained complete $BV$ light curves and used them to determine the photometric properties of the system and $U = 122 \pm 7$~yr. Lee et al.\ \cite{Lee+03jass} obtained four new timings and $U = 122.2 \pm 0.3$~yr, where the errorbar is anomalously small. Wolf et al.\ \cite{Wolf+06aa} presented nine new times of minimum and used these and published times to determine $U = 124.2 \pm 6.5$~yr. The most recent assessment of the apsidal motion period of \targ\ is by Claret et al.\ \cite{Claret+21aa} who included the TESS data to obtain $U = 116.2 \pm 3.9$~yr.

%%%%%%%%%%%%%%%%%%%%%%%%%%%%%%%%%%%%%%%%%%%%%%%%%%%%%%%%%%%%%%%%%%%%%%%%%%%%%%%%%%%%%%%%%%%%%%%%%%%%%%%%%%%%%%%%%%%%%%%%%%%%%%%%%%%%%%%%%%%%%%%%%%%%%

%\section*{\targfull}

\begin{table}[t]
\caption{\em Basic information on \targfull. The $BV$ magnitudes are each the mean of 93 individual measurements \cite{Hog+00aa}. \label{tab:info}}
\centering
\begin{tabular}{lll}
{\em Property}                            & {\em Value}                 & {\em Reference}                      \\[3pt]
Right ascension (J2000)                   & 03 59 44.68                 & \citenum{Gaia23aa}                   \\
Declination (J2000)                       & $+$48 09 04.5               & \citenum{Gaia23aa}                   \\
% Bright Star Catalogue                   & HR ----                     & \citenum{HoffleitJaschek91}          \\
Henry Draper designation                  & HD 24909                    & \citenum{CannonPickering18anhar}     \\
%\textit{Hipparcos} designation           & HIP 18662                   & \citenum{Hipparcos97}                \\
% \textit{Tycho} designation              & TYC 3331-1175-1             & \citenum{Hog+00aa}                   \\
\textit{Gaia} DR3 designation             & 246797724301643904          & \citenum{Gaia21aa}                   \\
\textit{Gaia} DR3 parallax                & $3.4478 \pm 0.0313$ mas     & \citenum{Gaia21aa}                   \\          % d = 287.5 +/- 2.6   RUWE = 0.999
TESS\ Input Catalog designation           & TIC 265767012               & \citenum{Stassun+19aj}               \\
$B$ magnitude                             & $7.778 \pm 0.013$           & \citenum{Hog+00aa}                   \\          % \cite{Henden+15aas} for APASS
$V$ magnitude                             & $7.733 \pm 0.012$           & \citenum{Hog+00aa}                   \\          % \cite{Hog+00aa} for Tycho
$J$ magnitude                             & $7.561 \pm 0.030$           & \citenum{Cutri+03book}               \\
$H$ magnitude                             & $7.591 \pm 0.031$           & \citenum{Cutri+03book}               \\
$K_s$ magnitude                           & $7.544 \pm 0.018$           & \citenum{Cutri+03book}               \\
Spectral type                             & B8~V + A6~V                 & \citenum{LacyFrueh85apj}             \\[3pt]
\end{tabular}
\end{table}

%%%%%%%%%%%%%%%%%%%%%%%%%%%%%%%%%%%%%%%%%%%%%%%%%%%%%%%%%%%%%%%%%%%%%%%%%%%%%%%%%%%%%%%%%%%%%%%%%%%%%%%%%%%%%%%%%%%%%%%%%%%%%%%%%%%%%%%%%%%%%%%%%%%%%

\section*{Photometric observations}

\begin{figure}[t] \centering \includegraphics[width=\textwidth]{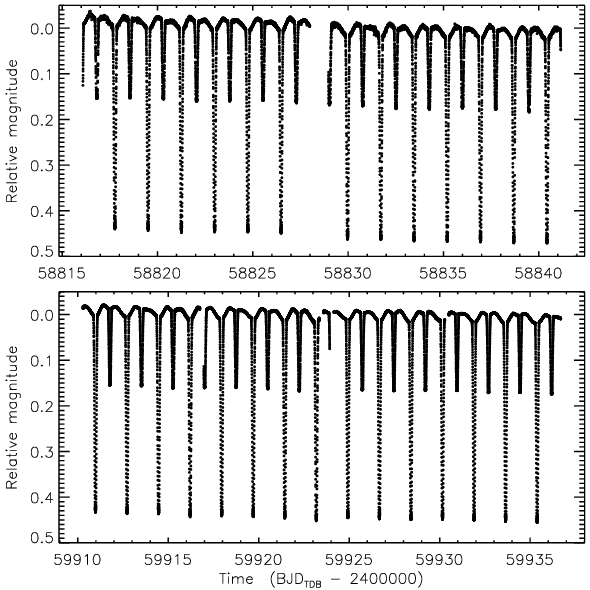} \\
\caption{\label{fig:time} TESS short-cadence SAP photometry of \targ\ from sectors 19 
(upper panel) and 59 (lower panel). The flux measurements have been converted to magnitude 
units then rectified to zero magnitude by subtraction of the median.} \end{figure}

% TESS-point Web Tool: Results Page for iq per
% The results for your object of interest can be seen below.
% 
% Results
% Name	RA	Dec	EclipLong	EclipLat	Sector	Camera	CCD	ColPix	RowPix	Sector Midpoint
% iq per	59.9362	48.1512	67.9718	26.9851	19	1	2	1625.0193	465.4433	December 2019
% iq per	59.9362	48.1512	67.9718	26.9851	59	1	2	1416.1745	728.1822	December 2022
% iq per	59.9362	48.1512	67.9718	26.9851	86	1	2	665.7092	789.174	December 2024

\targ\ has been observed in two sectors by TESS: sector 19 at cadences of 120~s and 1800~s; and sector 59 at cadences of 120~s and 200~s. Further observations are scheduled in sector 86 (2024 December). We downloaded both sets of 120-s cadence data from the NASA Mikulski Archive for Space Telescopes (MAST\footnote{\texttt{https://mast.stsci.edu/portal/Mashup/Clients/Mast/Portal.html}}) using the {\sc lightkurve} package \cite{Lightkurve18}. We adopted the simple aperture photometry (SAP) data from the SPOC data reduction \cite{Jenkins+16spie} with a quality flag of ``hard'', normalised them using {\sc lightkurve}, and converted them to differential magnitude.

The two light curves are shown in Fig.~\ref{fig:time}. That from sector 59 has a greater coverage and lower scatter, but the basic shape of the light variation is consistent between sectors. When the two light curves are overlaid on the same plot the eclipses do not quite line up. This is due to apsidal motion changing the relative times of eclipses in the 1069~d between the sectors. For the record, the numbers of datapoints are 17\,058 for sector 19 and 18\,367 for sector 59.

A query of the \gaia\ DR3 database\footnote{\texttt{https://vizier.cds.unistra.fr/viz-bin/VizieR-3?-source=I/355/gaiadr3}} returns a total of 142 sources within 2~arcmin of \targ, as expected due to the faint limiting magnitude of \gaia\ and the proximity of our target to the galactic plane. Aside from the nearby companion mentioned above, all of the stars are fainter by at least 5.5~mag in the $G$ band.
% This suggests that the TESS light curve of \targ\ will be contaminated by light from nearby stars at the level of only a few percent.

%%%%%%%%%%%%%%%%%%%%%%%%%%%%%%%%%%%%%%%%%%%%%%%%%%%%%%%%%%%%%%%%%%%%%%%%%%%%%%%%%%%%%%%%%%%%%%%%%%%%%%%%%%%%%%%%%%%%%%%%%%%%%%%%%%%%%%%%%%%%%%%%%%%%%

\section*{Preliminary light curve analysis}

We performed a first analysis of the data using version 43 of the {\sc jktebop}\footnote{\texttt{http://www.astro.keele.ac.uk/jkt/codes/jktebop.html}} code \cite{Me++04mn2,Me13aa}. We concentrated on the data from sector 59, as these are of higher quality that those from sector 19 and we do not want to combine the sectors due to the motion of the argument of periastron. This work confirmed that the fractional radius of star~A ($r_{\rm A} = {R_{\rm A}}/{a}$ where $R_{\rm A}$ is its radius and $a$ the semimajor axis of the relative orbit) is close to the limits of applicability of {\sc jktebop} and thus it was advisable to use a more sophisticated code for the final analysis.

We therefore fitted the sector 59 light curve using {\sc jktebop} only to determine the orbital ephemeris and the coefficients of a polynomial to normalise the brightness of the system to zero differential magnitude. We then subtracted the polynomial, converted the times of observation to orbital phase, and combined them into 1000 phase bins. The resulting phase-binned data were then suitable for the next step in the analysis.

%%%%%%%%%%%%%%%%%%%%%%%%%%%%%%%%%%%%%%%%%%%%%%%%%%%%%%%%%%%%%%%%%%%%%%%%%%%%%%%%%%%%%%%%%%%%%%%%%%%%%%%%%%%%%%%%%%%%%%%%%%%%%%%%%%%%%%%%%%%%%%%%%%%%%

\section*{Analysis with the Wilson-Devinney code}

The main analysis of the light curve was undertaken using the Wilson-Devinney (WD) code \cite{WilsonDevinney71apj,Wilson79apj}, which uses Roche geometry to represent the shapes of the stars. We used the 2004 version of the code ({\sc wd2004}), driven by the {\sc jktwd} wrapper \cite{Me+11mn}, to fit the 1000-point phase-binned light curve. Following our usual practice, we first describe the adopted solution of the light curve and then discuss the alternative approaches which comprise our error analysis. The parameters used in {\sc wd2004} are described in the WD code user manual \cite{WilsonVanhamme04}.

The fitted parameters in the adopted solution were: the potentials of the two stars; the orbital inclination; the orbital eccentricity; the argument of periastron; a phase offset; one limb darkening (LD) coefficient per star; the light contribution of star~A; the effective temperature (\Teff) of star~B; and the amount of third light. For LD we used the logarithmic law \cite{KlinglesmithSobieski70aj} with the nonlinear coefficient fixed \cite{Me23obs2} to theoretical values from Van Hamme \cite{Vanhamme93aj}. Albedo values and gravity darkening exponents were fixed to 1.0 (suitable for radiative atmospheres), synchronous rotation was assumed, the simple model of reflection was used, and the mass ratio was fixed to the spectroscopic value from LF85.

Our adopted solution differs from our usual approach in that we have chosen as default to operate the WD code in {\sc mode}=2, where the \Teff\ values and light contributions of the stars are linked. Our initial experiments using {\sc mode}=0 gave similar fits but for albedo values significantly greater than unity, a situation we have noticed several times in the past \cite{Me20obs,Me+11mn,Pavlovski++18mn,Pavlovski+23aa}. This problem does not occur in WD {\sc mode}=2. Table~\ref{tab:wd} contains the parameters of our adopted fit. 

\begin{table} \centering
\caption{\em Summary of the parameters for the {\sc wd2004} solution of the TESS light curve of \targ. Uncertainties
are only quoted when they have been assessed by comparison between a full set of alternative solutions. \label{tab:wd}}
% \addtolength{\tabcolsep}{-2pt}
\begin{tabular}{lcc}
{\em Parameter}                           & {\em Star A}          & {\em Star B}          \\[3pt]           % & Symbol
{\it Control parameters:} \\
{\sc wd2004} operation mode               & \multicolumn{2}{c}{2}                         \\                % & {\sc mode}
Treatment of reflection                   & \multicolumn{2}{c}{1}                         \\                % & {\sc mref}
Number of reflections                     & \multicolumn{2}{c}{1}                         \\                % & {\sc nref}
Limb darkening law                        & \multicolumn{2}{c}{2 (logarithmic)}           \\                % & {\sc ld}
Numerical grid size (normal)              & \multicolumn{2}{c}{60}                        \\                % & {\sc n1, n2}
Numerical grid size (coarse)              & \multicolumn{2}{c}{60}                        \\[3pt]           % & {\sc n1l, n2l}
{\it Fixed parameters:} \\
% Orbital period (d)                      & \multicolumn{2}{c}{1.6697739}                 \\                % & {\sc period}
% Primary eclipse time (BJD$_{\rm TDB}$)  & \multicolumn{2}{c}{2458366.46953}             \\                % & {\sc hjd0}
Phase shift                               & \multicolumn{2}{c}{0.0}                       \\                % & {\sc pshift}
Mass ratio                                & \multicolumn{2}{c}{0.493}                     \\                % & {\sc rm}
Rotation rates                            & 1.0                   & 1.0                   \\                % & {\sc f1, f2}
Bolometric albedos                        & 1.0                   & 1.0                   \\                % & {\sc alb1, alb2}
Gravity darkening                         & 1.0                   & 1.0                   \\                % & {\sc gr1, gr2}
\Teff\ value of star~A (K)                & 12300                 &                       \\                % & {\sc tavh}
Bolometric linear LD coefficient          & 0.7321                & 0.6565                \\                % & {\sc xbol1, xbol2}
Bolometric logarithmic LD coefficient     & 0.0714                & 0.2421                \\                % & {\sc xbol1, xbol2}
Passband logarithmic LD coefficient       & 0.2354                & 0.2849                \\[3pt]           % &
{\it Fitted parameters:} \\
Potential                                 & $4.838 \pm 0.019$     & $4.814 \pm 0.050$     \\                % & {\sc phsv, pcsv}
Orbital inclination (\degr)               & \multicolumn{2}{c}{$88.63 \pm 0.52$}          \\                % & {\sc xincl}
Orbital eccentricity                      & \multicolumn{2}{c}{$0.0677 \pm 0.0022$}       \\                % & {\sc e}
Argument of periastron (\degr)            & \multicolumn{2}{c}{$200.0 \pm 5.2$}           \\                % & {\sc perr0}
\Teff\ value of star~B (K)                &                       & $8180 \pm 360$        \\                % & {\sc tavc}
Light contribution                        & $10.30 \pm 0.18$      &                       \\                % & {\sc hlum}, {\sc clum}
Passband linear LD coefficient            & $0.486 \pm 0.067$     & $0.38 \pm 0.17$       \\                % & {\sc ldu}
Third light                               & \multicolumn{2}{c}{$0.054 \pm 0.014$}         \\                % & {\sc el3}
{\it Derived parameters:} \\
Fractional radii                          & $0.2336 \pm 0.0009$   & $0.1418 \pm 0.0014$   \\[10pt]          % &
% Light ratio                             & \multicolumn{2}{c}{$0.754 \pm 0.036$}         \\                % &
\end{tabular}
\end{table}

\begin{figure}[t] \centering \includegraphics[width=\textwidth]{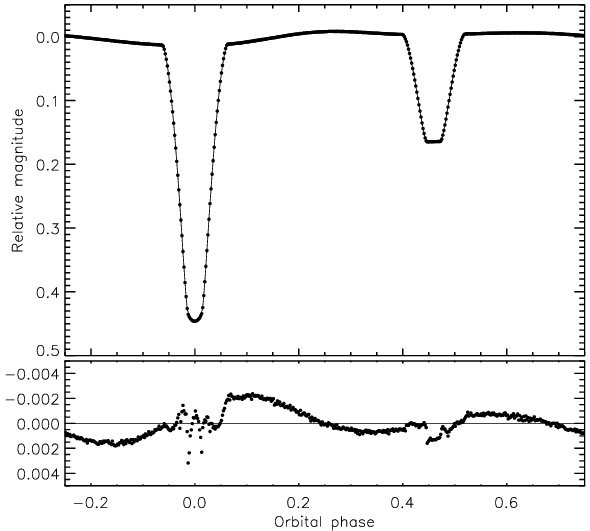} \\
\caption{\label{fig:phase} Best fit to the binned light curve of \targ\ using {\sc wd2004}.
The phase-binned data are shown using open circles and the best fit with a continuous line.
The residuals are shown on an enlarged scale in the lower panel.} \end{figure}

Fig.~\ref{fig:phase} shows the fit of the model to the data, which is imperfect. We tried all possible options available in the WD code to improve the fit (see our work on V1388~Ori; ref.\ \cite{Me22obs4}) but were unable to do so. There is an increased scatter through the primary eclipse, which is due to the numerical resolution of the {\sc wd2004} code. There is a sinusoidal trend in the residuals at twice the orbital frequency; its shape is not consistent with any of the ellipsoidal, reflection or Doppler beaming effects \cite{MorrisNaftilan93apj,Zucker++07apj}. A similar but not identical trend has previously been seen for the dEBs $\zeta$~Phe \cite{Me20obs} and KIC~9851944 \cite{Jennings+24mn}. Possible causes include the assumption of point masses in the Roche model, and thus neglect of the mass of the envelopes of the stars, and the brightness changes due to pulsations (see below). The relatively poor fit during secondary eclipse is an artefact of the sinusoidal trend, which causes incorrect normalisation of the light curve in the region of the secondary eclipse and a slight deformation of the eclipse to obtain the overall best least-squares fit. In our analysis below we make the assertion that our best fit -- whilst not a good fit -- nevertheless yields reliable parameters which can be used to determine the physical properties of the component stars; the reader is free to disagree if they wish.

We determined the errorbars of the fitted parameters in the {\sc wd2004} analysis by considering a range of possible choices in delineating the adopted solution. The scatter of the data is small, and the residuals versus the best fit are significantly larger, so the contribution of Poisson noise is negligible. Instead we ran a set of alternative solutions varying one or more control parameters or input physics from the adopted solution. These alternative solutions included: use of a light curve binned into 500 phase points instead of 1000; numerical precisions of 40 and 55--59 instead of the maximum 60; a different approach to polynomial normalisation of the data before phase-binning; changing the spectroscopic mass ratio by its uncertainty; changing the rotation rates by $\pm$0.1, changing the gravity darkening exponents by $\pm$0.1; changing the albedos by $\pm$0.1; using the square-root LD law; fixing all LD coefficients at the theoretically-predicted values; and using the Cousins $I$ band instead of $R$ band as a surrogate for the TESS passband.

The result of this process was a large set of alternative parameter values. The differences for each parameter versus the adopted solution were calculated then added in quadrature to obtain the final uncertainty for that parameter. These are the errorbars reported in Table~\ref{tab:wd}. We also give a breakdown in Table~\ref{tab:rerror} of the errorbars for the fractional radii, as these are the most important result from the WD analysis. The final uncertainties in the fractional radii are 0.4\% for star~A and 1.0\% for star~B. The \Teff\ of star~B from this analysis is not reliable because the TESS passband is not an available option for the {\sc wd2004} code. We also obtained a fit to the TESS sector 19 data, finding it to be consistent with that for the sector 59 data but with a different small trend in the residuals. We postpone further analysis of these trends to a future work.

\begin{table} \centering
\caption{\em Changes in the measured fractional radii of the stars due to differing model choices. 
Each is expressed as the percentage change versus the value of the parameter. \label{tab:rerror}}
\begin{tabular}{lrr}
Model choice                                      &       \mc{~Effect (\%)}     \\
                                                  & $r_{\rm A}$~ &  $r_{\rm B}$~ \\[3pt]
Binning into 500 phase bins instead of 1000                 &$-$0.06 &    0.15 \\
Setting the numerical precision to {\sc n1}$=${\sc n2}$=$40 &   0.07 & $-$0.32 \\
Different polynomial normalisation                          &   0.16 &    0.19 \\
Changing mass ratio                                         &   0.00 & $-$0.07 \\
Changing rotation rates by $\pm$0.1                         &$-$0.09 &    0.05 \\
Changing gravity darkening by $\pm$0.1                      &$-$0.15 &    0.22 \\
Changing the albedos by $\pm$0.1                            &   0.02 &    0.03 \\
Using the square-root limb darkening law                    &$-$0.02 & $-$0.05 \\
Fixing limb darkening coefficients                          &   0.00 &    0.41 \\
Using the Cousins $I$-band                                  &   0.30 & $-$0.72 \\
% Using the detailed treatment of reflection                &   0.01 &    0.00 \\
% Detailed treatment of reflection with two reflections     &   0.01 & $-$0.00 \\
% Fitting for phase shift                                   &   0.00 &    0.00 \\
\end{tabular}
\end{table}

%%%%%%%%%%%%%%%%%%%%%%%%%%%%%%%%%%%%%%%%%%%%%%%%%%%%%%%%%%%%%%%%%%%%%%%%%%%%%%%%%%%%%%%%%%%%%%%%%%%%%%%%%%%%%%%%%%%%%%%%%%%%%%%%%%%%%%%%%%%%%%%%%%%%%%

\section*{Radial velocity analysis}

LF85 presented 20 medium-resolution spectra, and measured 20 RVs for star~A and 16 for star~B using cross-correlation. We digitised the data and modelled the RVs with the {\sc jktebop} code to determine the velocity amplitudes of the two stars. A time of minimum close to the mean time of the spectra was chosen\footnote{HJD 2444926.749 from table~5 of LF85} and a shift in orbital phase was included as a fitted parameter. We also fitted for the velocity amplitudes ($K_{\rm A}$ and $K_{\rm B}$) and systemic velocities ($V_{\rm \gamma,A}$ and $V_{\rm \gamma,B}$) of the stars. Due to the apsidal motion in the system, we fitted for the argument of periastron ($\omega$) but fixed the eccentricity at the photometric value (Table~\ref{tab:wd}). Alternative solutions with eccentricity fitted or with $V_{\rm \gamma,A} = V_{\rm \gamma,B}$ gave results which differed by much less than the uncertainties, indicating that the orbital solutions are robust. We followed LF85 by allocating half weight to two spectra with a lower count rate, and iteratively adjusted the errorbars of the RVs of each star to obtain a reduced \chisq\ of 1.0.

% Vsys same, e fixed:   K1 = 101.95 \pm 0.65   K2 = 206.16 \pm 1.17   Vsys1 = 0.57 \pm 0.41                           e = 0.0677              w = 73.1 \pm 4.8     (MC errorbars)
% Vsys same, e fitted:  K1 = 102.04 \pm 0.65   K2 = 206.32 \pm 1.15   Vsys1 = 0.56 \pm 0.39                           e = 0.0715 \pm 0.0022   w = 72.3 \pm 4.6     (MC errorbars)
% Vsys diff, e fixed:   K1 = 101.95 \pm 0.65   K2 = 206.16 \pm 1.20   Vsys1 = 0.55 \pm 0.47   Vsys2 = 0.65 \pm 0.88   e = 0.0677              w = 73.0 \pm 5.0     (MC errorbars) ** ADOPTED **
% Vsys diff, e fitted:  K1 = 102.14 \pm 0.61   K2 = 206.53 \pm 1.21   Vsys1 = 0.17 \pm 0.56   Vsys2 = 1.68 \pm 1.13   e = 0.0758 \pm 0.0061   w = 70.0 \pm 4.9     (MC errorbars)
  
\begin{figure}[t] \centering \includegraphics[width=\textwidth]{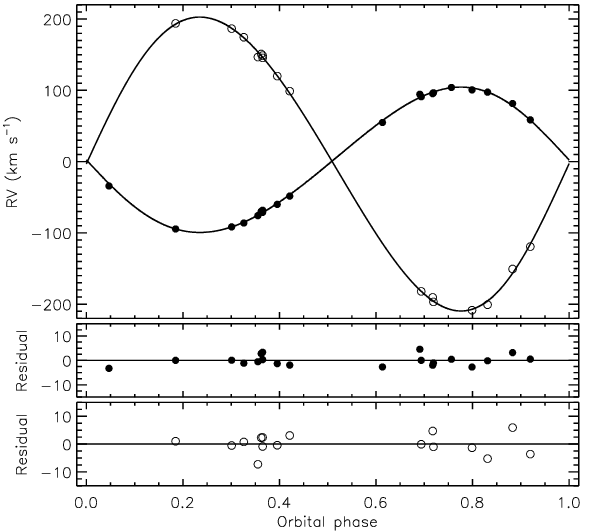} \\
\caption{\label{fig:rv} RVs of \targ\ from LF85 (filled circles for star~A and open 
circles for star~B), compared to the best fit from {\sc jktebop} (solid lines). The 
residuals are given in the lower panels separately for the two components.} \end{figure}

The fitted orbits are shown in Fig.~\ref{fig:rv}. The parameters of the fit are $K_{\rm A} = 101.95 \pm 0.65$\kms, $K_{\rm B} = 206.2 \pm 1.2$\kms, $V_{\rm \gamma,A} = 0.55 \pm 0.47$\kms, $V_{\rm \gamma,B} = 0.65 \pm 0.88$\kms\ and $\omega = 73.0 \pm 5.0$. These are all in good agreement with the values found by LF85, but have smaller errorbars because we fixed the eccentricity to a known value instead of fitting it separately for the two stars. 

%%%%%%%%%%%%%%%%%%%%%%%%%%%%%%%%%%%%%%%%%%%%%%%%%%%%%%%%%%%%%%%%%%%%%%%%%%%%%%%%%%%%%%%%%%%%%%%%%%%%%%%%%%%%%%%%%%%%%%%%%%%%%%%%%%%%%%%%%%%%%%%%%%%%%%

\section*{Physical properties and distance to \targ}

\begin{table} \centering
\caption{\em Physical properties of \targ\ defined using the nominal solar units given by 
IAU 2015 Resolution B3 (ref.\ \cite{Prsa+16aj}). The \Teff\ values are from LF85. \label{tab:absdim}}
\begin{tabular}{lr@{\,$\pm$\,}lr@{\,$\pm$\,}l}
{\em Parameter}        & \multicolumn{2}{c}{\em Star A} & \multicolumn{2}{c}{\em Star B}    \\[3pt]
Mass ratio   $M_{\rm B}/M_{\rm A}$          & \multicolumn{4}{c}{$0.4944 \pm 0.0043$}       \\
Semimajor axis of relative orbit (\Rsunnom) & \multicolumn{4}{c}{$10.598 \pm 0.047$}        \\
Mass (\Msunnom)                             &  3.516  & 0.050       &  1.738  & 0.023       \\
Radius (\Rsunnom)                           &  2.476  & 0.015       &  1.503  & 0.016       \\
Surface gravity ($\log$[cgs])               &  4.1967 & 0.0042      &  4.3244 & 0.0090      \\
Density ($\!\!$\rhosun)                     &  0.2317 & 0.0028      &  0.512  & 0.015       \\
Synchronous rotational velocity ($\!\!$\kms)& 71.84   & 0.42        & 43.61   & 0.47        \\
Effective temperature (K)                   & 12300   & 170         &  7675   & 100         \\
Luminosity $\log(L/\Lsunnom)$               &   2.102 & 0.025       &  0.849  & 0.022       \\
$M_{\rm bol}$ (mag)                         &$-$0.514 & 0.061       &  2.618  & 0.056       \\
Interstellar reddening \EBV\ (mag)          & \multicolumn{4}{c}{$0.12 \pm 0.03$}			\\
Distance (pc)                               & \multicolumn{4}{c}{$278.1 \pm 3.8$}           \\[3pt]
\end{tabular}
\end{table}

% IDL> print, [0.050,0.023,0.015,0.016]/[3.516,1.738,2.76,1.503]*100
%       1.42207      1.32336     0.543478      1.06454

We determined the physical properties of \targ\ from the results of the {\sc wd2004} code and RV analyses given above. For this we used the {\sc jktabsdim} code \cite{Me++05aa}. The results are given in Table~\ref{tab:absdim}. The masses of the stars are measured to 1.5\% precision, and the radii to 0.5\% (star~A) and 1.1\% (star~B) assuming the {\sc wd2004} fit is good enough to give reliable parameters. Our results are in excellent agreement with those from LF85, highlighting the robustness of such information for totally-eclipsing binaries (and our use of the same RV data). We also retain the \Teff\ values given by LF85, which were based on photometric colour indices for the individual stars. The pseudo-synchronous rotational velocities are consistent with the values measured by LF85.

We estimated the distance to the system using our measured radii, the apparent magnitudes in Table\,\ref{tab:info}, and bolometric corrections from Girardi et al.\ \cite{Girardi+02aa}. Imposing an interstellar reddening of $\EBV = 0.12 \pm 0.03$~mag to bring the $BV$ and $JHK_s$-band distances into agreement, we obtained a distance of $278.1 \pm 3.8$~pc. This is a reasonable match to the distance of $287.5 \pm 2.6$~pc from \gaia\ DR3 parallax of \targ, supporting the reliability of the \Teff\ values from LF85. 

We compared the measured properties of the components of \targ\ to the predictions of the {\sc parsec} theoretical stellar evolutionary models \cite{Bressan+12mn}. The large difference between the two stars means they are a high-quality test of theoretical predictions, and in this case the test is failed. We can obtain a good match to the masses and radii of the stars for a metal abundance of $Z = 0.017$ and an age of 80~Myr, but the predicted \Teff\ values are too large ($\sim$16000 and $\sim$8200, respectively). A higher or lower metal abundance gives a poorer fit to the radii. A metal abundance of $Z=0.030$ and an age of 45~Myr can match the radius of star~A and the \Teff\ of star~B, but not the radius of star~B and the \Teff\ of star~A. A spectroscopic analysis to confirm the \Teff\ values and obtain a metallicity measurement for the stars would be very helpful in exploring this discrepancy further.

% Z=0.014 and 90Myr fits R1
% Z=0.017 and 80Myr fits R1 and R2
% Z=0.020 and 70Myr fits R1
% Z=0.030 and 45Myr fits R1 and T2

%%%%%%%%%%%%%%%%%%%%%%%%%%%%%%%%%%%%%%%%%%%%%%%%%%%%%%%%%%%%%%%%%%%%%%%%%%%%%%%%%%%%%%%%%%%%%%%%%%%%%%%%%%%%%%%%%%%%%%%%%%%%%%%%%%%%%%%%%%%%%%%%%%%%%

\section*{Pulsation analysis}

\begin{sidewaysfigure} \centering
\includegraphics[width=\textwidth]{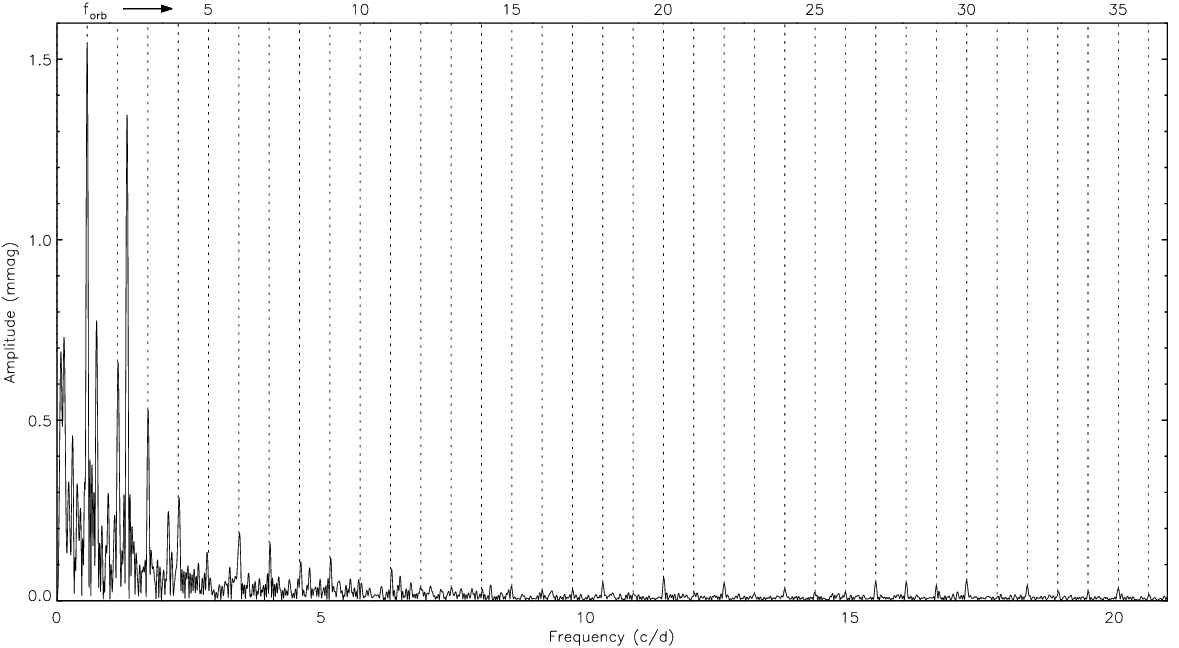}
\caption{\label{fig:freq} Frequency spectrum of the TESS light curve  
of \targ. Multiples of the orbital frequency are shown along the top.} 
\end{sidewaysfigure}

An increasing number of dEBs are known to harbour pulsating stars \cite{Chen+22apjs,EzeHandler24apjs}. We performed a search for pulsations in \targ\ using the residuals of the {\sc jktebop} fit to the TESS sector 59 light curve. An amplitude spectrum was calculated using version 1.2.0 of the {\sc period04} code \cite{LenzBreger05coast} and is shown in Fig.~\ref{fig:freq}. We find a large number of small signals at multiples of the orbital frequency, which is expected due to the trend with orbital phase seen in the residuals from the {\sc jktebop} and {\sc wd2004} fits (e.g.\ Fig.~\ref{fig:phase}). No significant frequencies were found from 22\cd\ up to the Nyquist frequency of 360\cd.

% 0.749775351 0.000291738154	snr: 2.65813
% 1.32725384  0.000215412941	snr: 6.24727   amp=0.00134916169
% 2.10921652  0.000130265604	snr: 1.89213

There are three frequencies which are not at multiples of the orbital frequency, at 0.75, 1.33 and 2.11\cd. 
%These are clearly related as the two higher frequencies are twice and three times the lowest frequency. 
These are likely related as the highest frequency is approximately the sum of the two lower frequencies.
Of these, only the 1.33\cd\ frequency is significant \cite{BaranKoen21aca}, with a signal-to-noise ratio of 6.2 and an amplitude of 1.3~mmag (calculated using {\sc period04}). Based on this detected frequency and the masses of the stars, we conclude that star~A is a slowly-pulsating B-star \cite{Waelkens91aa,Pedersen22apj}. It thus joins the small but increasing sample of such stars in dEBs \cite{MeBowman22mn}. The prospects for asteroseismology of this star are poor as only one significant pulsation has been detected with a frequency not corresponding to a multiple of the orbital frequency.

%%%%%%%%%%%%%%%%%%%%%%%%%%%%%%%%%%%%%%%%%%%%%%%%%%%%%%%%%%%%%%%%%%%%%%%%%%%%%%%%%%%%%%%%%%%%%%%%%%%%%%%%%%%%%%%%%%%%%%%%%%%%%%%%%%%%%%%%%%%%%%%%%%%%%%

\section*{Conclusion}

\targ\ is a dEB containing a 3.5\Msun\ B-star and a 1.7\Msun\ A-star in an orbit of short period (1.744~d) which is eccentric and shows apsidal motion. We have used new light curves from TESS and published RVs from LF85 to determine the masses and radii of the component stars to high precision (0.5--1.5\%). The significant differences between the two stars makes \targ\ a good test of stellar theory. The masses, radii and \Teff\ values of the stars cannot be fitted for a single age and metallicity using the {\sc parsec} models. A more extensive analysis could be performed by obtaining spectroscopic \Teff\ and metallicity measurements and by including in the analysis internal structure constants measured from the apsidal motion.

Our fit to the TESS light curve is imperfect, with a roughly sinusoidal residual versus orbital phase and a slight mismatch of the depth of the secondary eclipse. However, the total eclipses mean the radii of the stars can still be measured reliably from the times of the contact points. A frequency spectrum of the residuals of the fit has many peaks at multiples of the orbital frequency, as expected due to the residuals versus the best fit. It also shows one significant peak at 1.33\cd, away from multiples of the orbital frequency, 
%and smaller peaks at 0.5 and 1.5 times this frequency
and less significant peaks at 0.75\cd\ and 2.11\cd. These results are consistent with star~A being a slowly-pulsating B-star.

%%%%%%%%%%%%%%%%%%%%%%%%%%%%%%%%%%%%%%%%%%%%%%%%%%%%%%%%%%%%%%%%%%%%%%%%%%%%%%%%%%%%%%%%%%%%%%%%%%%%%%%%%%%%%%%%%%%%%%%%%%%%%%%%%%%%%%%%%%%%%%%%%%%%%%%

\section*{Acknowledgements}

%We thank the anonymous referee for a positive and extraordinarily prompt report.
This paper includes data collected by the TESS\ mission and obtained from the MAST data archive at the Space Telescope Science Institute (STScI). Funding for the TESS\ mission is provided by the NASA's Science Mission Directorate. STScI is operated by the Association of Universities for Research in Astronomy, Inc., under NASA contract NAS 5–26555.
This work has made use of data from the European Space Agency (ESA) mission {\it Gaia}\footnote{\texttt{https://www.cosmos.esa.int/gaia}}, processed by the {\it Gaia} Data Processing and Analysis Consortium (DPAC\footnote{\texttt{https://www.cosmos.esa.int/web/gaia/dpac/consortium}}). Funding for the DPAC has been provided by national institutions, in particular the institutions participating in the {\it Gaia} Multilateral Agreement.
The following resources were used in the course of this work: the NASA Astrophysics Data System; the SIMBAD database operated at CDS, Strasbourg, France; and the ar$\chi$iv scientific paper preprint service operated by Cornell University.

%%%%%%%%%%%%%%%%%%%%%%%%%%%%%%%%%%%%%%%%%%%%%%%%%%%%%%%%%%%%%%%%%%%%%%%%%%%%%%%%%%%%%%%%%%%%%%%%%%%%%%%%%%%%%%%%%%%%%%%%%%%%%%%%%%%%%%%%%%%%%%%%%%%%%

% \bibliographystyle{obsmaga}
% \bibliography{jkt}

%%%%%%%%%%%%%%%%%%%%%%%%%%%%%%%%%%%%%%%%%%%%%%%%%%%%%%%%%%%%%%%%%%%%%%%%%%%%%%%%%%%%%%%%%%%%%%%%%%%%%%%%%%%%%%%%%%%%%%%%%%%%%%%%%%%%%%%%%%%%%%%%%%%%%
\end{document}